\documentclass[%
 reprint,
 superscriptaddress,
 amsmath,amssymb,
 aps,
 pre,
 floatfix,
]{revtex4-1}

\usepackage{graphicx} % Include figure files
\usepackage{xcolor}   % Add coloring options
\usepackage{dcolumn}  % Align table columns on decimal point
\usepackage{bm}       % bold math
\usepackage{hyperref} % add hypertext capabilities
\hypersetup{          % set up additional options for hyperref
 pdftitle={Geometrical Interpretation of Dynamical Phase Transitions in Boundary Driven Systems   },
 pdfauthor={Ohad Shpielberg},
 pdfsubject={},
 pdfkeywords={},
 unicode=true,
 breaklinks=true,
 linkcolor=blue, 
 citecolor=[rgb]{0.0,0.7,0.1},
 pdfborder={0 0 0},
 pdfborderstyle={},
 colorlinks=true,
 bookmarksnumbered=true,
 bookmarksopen=true,
 bookmarksopenlevel=1,
}

\usepackage[all]{xy}                        % typesetting graphs & diagrams
\SelectTips{cm}{10}                         % better tips
\usepackage[abbreviations,xlatin]{foreign}  % proper typeset of abbreviations & Latin
\defasforeign[aesthetic]{\ae{}sthetic}      % use Latin æsc for aesthetic

\usepackage[normalem]{ulem}                 % strikeout
\usepackage{comment}                        % debug
\begin{document}

% \preprint{APS/123-QED}

\title{Geometrical Interpretation of Dynamical Phase Transitions in Boundary Driven Systems}
%
% \thanks{A footnote to the article title} 

\author{Ohad Shpielberg}
\email{ohad.shpielberg@lpt.ens.fr}
\affiliation{Laboratoire de Physique Th\'{e}orique de l'\'{E}cole Normale Sup\'{e}rieure de Paris, CNRS, ENS  \& PSL Research University,UPMC  \& Sorbonne Universit\'{e}s, 75005 Paris, France.}
%\affiliation{Physics Department, Technion--Israel Institute of Technology, 3200003 Haifa, Israel}

%

\date{\today} % It is always \today, today,
%             %  but any date may be explicitly specified

% ======================================================================================== %

\begin{abstract}

Dynamical phase transitions are defined as   non-analytic points of the large deviation function of current fluctuations. We show that for boundary driven systems,  many dynamical phase transitions can be identified using the geometrical structure of an effective potential of a Hamiltonian, recovered from the macroscopic fluctuation theory description.
Using this method we identify new dynamical phase transitions that could not be recovered using existing perturbative methods.  
Moreover, using the Hamiltonian picture,  an experimental scheme is suggested to demonstrate an analog of dynamical phase transitions in linear, rather than exponential, time.

\end{abstract}

% \pacs{Valid PACS appear here}% PACS, the Physics and Astronomy
\maketitle

% \tableofcontents

% ======================================================================================== %

\section{Introduction
\label{sec:introduction}}

The study of phase transitions spans across all branches of physics \citep{Wilson74,Yeomans,Kosterlitz73,Sachdev}. Thermodynamic phase transitions in equilibrium systems have been studied extensively \citep{Callen,LandauLifshiftz1980StatisticalPt1}. However, for systems driven out-of-equilibrium, even simple ideas valid in equilibrium seem to be violated \cite{Derrida2007}. As a prominent example, we note the Peierls argument:  There are no phase transitions in equilibrium $1d$ systems with short-range interactions \citep{LandauLifshiftz1980StatisticalPt1}. This argument  breaks down for out-of-equilibrium systems \citep{Mallick2015,Aminov2014,Bertini2010a,Bodineau2005,Appert-Rolland2008,Tizon16a,Tizon16b,Zarfaty2015a}. While out-of-equilibrium systems allow for a richer set of effects,  a theory comparable to statistical mechanics is lacking. 

A major advancement in the understanding of out-of-equilibrium systems is the recent formulation of the macroscopic fluctuation theory (MFT) \citep{Bertini2015,Bertini2001}. The MFT offers a hydrodynamic description of steady state fluctuations in diffusive systems. It was used to characterize steady state density correlations \citep{Bertini2009,Bertini2007a}, identify fluctuation induced forces \citep{Aminov15}, find Clausius inequalities \citep{Bertini2013}, follow the statistics of single file diffusion \citep{Krapivsky2015} and many more  \cite{Akkermans2013,Bodineau2010a,Bodineau2008a,Agranov2016,Hirschberg2015,Agranov2017_arxiv,Hurtado2013a}. With the help of the MFT, it is conceivable that a classification of phase transitions in out-of-equilibrium diffusive systems  can be pursued. 

In this paper, we focus on  current fluctuations in boundary-driven diffusive systems within the framework of the MFT.   The study of current fluctuations deals with finding the probability $P_t(Q)$  to observe an atypical charge  transfer $Q$  at the long time limit $t$  \footnote{We could also consider continuous observables instead of discrete charge, e.g. heat.}. Here we restrict the discussion to $1d$ systems only. We assume throughout the text that a large deviation principle applies, namely   
\begin{equation}
P_t(Q) \sim \exp \left(-t \, \Phi\left(J=Q/t\right)\right),
\end{equation}
where $\Phi\left(J\right)$ is the large deviation function (LDF) of $J$, the mean current in the system. Obtaining an exact expression for $\Phi$ is not a trivial task both analytically \citep{Bernard2012,Meerson2016,Derrida2016,Sadhu2015,Touchette2009,DeGier2011,Bodineau2010,Agranov2016} and numerically \cite{Giardina2006,Hurtado2009,Nemoto2017}. In boundary driven diffusive systems, the MFT allows to write the LDF formally. Finding $\Phi\left(J\right)$ reduces to finding the optimal fluctuation of the density profile responsible for the atypical current.   Finding this optimal fluctuation,  is inherently hard. 
In \cite{Bodineau2004}, it was suggested that the optimal fluctuation  is time-independent (except for a negligible transient time). This conjecture, known as the additivity principle (AP), was shown to be exact for several boundary driven systems \citep{Bodineau2004,Bertini2006,Hurtado2008,Shpielberg2016}  and is believed to be always valid for currents sufficiently close to the steady state current.  In fact, a violation of the AP was found only recently for boundary driven systems  \cite{Shpielberg2016,Shpielberg2017a}. Obtaining the large deviation function under the AP assumption boils down to solving boundary valued non-linear differential equations. This allows for the possibility of multiple solutions as was demonstrated in \cite{Bertini2005,Baek2016b}. Multiple optimal solutions may lead to non-analytic points in the large deviation function \cite{Bertini2005,Imparato2009,Appert-Rolland2008}. Transitions between optimal solutions as  a function of $J$  (AP violations included) go under the name of dynamical phase transitions (DPT). It is usually hard to find analytically all possible solutions. Moreover, it is appealing to  be able to predict the occurrence of  DPTs from simple arguments. This goal is pursued here using a mapping to a one-body Lagrangian mechanics, where DPTs are identified as non-analytic points in a minimization of an action. Moreover, an experimental setup is proposed to observe an analog of DPTs in linear, rather then exponential time.    

This paper is organized as follows. In section \ref{sec:background} we briefly recapitulate the MFT and the AP conjecture, as well as the 
analogy to Lagrangian mechanics. In section \ref{sec:Dynamical Phase transitions} a few relevant models are considered to  demonstrate the required essentials for a  
DPT. In section \ref{sec:weak fields} we generalize the geometrical method to boundary driven processes with a weak field. In section \ref{sec:summary} we summarize the results and  discuss  future directions. Moreover, an experimental setup is proposed in which the analog of  DPTs can be can be directly observed.

% ======================================================================================== %

\section{Theoretical Background
\label{sec:background}}

This section is devoted to summarizing the main points of the MFT leading to the AP conjecture. We will then present the mapping to Lagrangian mechanics, which will later prove useful to identifying DPTs. 

Consider a lattice gas in a $1d$ system of $L$ sites with diffusive dynamics. In the fluctuating hydrodynamic approach \citep{Tailleur07,Imparato2009,Jordan2004},   we replace the microscopic space and time coordinates $i\in 1,...,L$ and $t$ with hydrodynamic coordinates $x=i/L \in \left[0,1\right]$ and  $t' = t/L^2 $. The relevant macroscopic variables are the particle  and current densities $\rho\left(x,t'\right),j\left(x,t'\right)$ \cite{Derrida2007},  related through the continuity equation 
\begin{equation}
\partial_\tau \rho\left(x,t'\right) = -\partial_x j 	 \left(x,t'\right).
\label{eq:continuity equation}
\end{equation}
Connecting our system to two reservoirs at the boundaries $x=0,1$ with fixed densities $\rho_l,\rho_r$ correspondingly, drives it out of equilibrium. The steady state current $J_S$ and steady state density profile $\rho_S$ of such a diffusive system obey Fick's law 
\begin{equation}
J_S  = -D\left(\rho_S\right) \partial_x \rho_S. \label{eq:Fick}
\end{equation}
Using \eqref{eq:Fick} in \eqref{eq:continuity equation} identifies $D$ as the diffusion parameter of the (steady state) diffusion equation. 
Assuming that the current density $j\left(x,t'\right)$ can be described by small fluctuations around the steady state provides a description of the dynamics. This amounts to writing 
\begin{equation}
j\left(x,t'\right) = -D\left(\rho\left(x,t'\right)\right) \partial_x \rho\left(x,t'\right) + \sqrt{\frac{\sigma(\rho\left(x,t'\right))}{L}}\xi\left(x,t'\right), \label{eq:fluc hydro}
\end{equation}
with $\xi$ a white noise in $\left( x,t' \right)$, and $\sigma(\rho)$, the conductivity \citep{Derrida2007}, characterizes   the fluctuations. %\red{The ratio of $D$ and $\sigma$ is determined by the Einstein relation \citep{Derrida2007}.}

Using the Martin-Siggia-Rose procedure \citep{Martin1973} for \eqref{eq:fluc hydro},
one finds that the probability to observe $\lbrace\rho,j\rbrace$ in time and space is given by
\begin{equation}
\mathcal{P}_t \left(\lbrace\rho,j\rbrace\right) \sim \exp \left( -L\int^1 _0 dx \int^{t/L^2}_0 d t'    \,    \mathcal{L}     \right), \label{eq:fund MFT}
\end{equation}
where $\mathcal{L}=\frac{1}{2\sigma\left(\rho\right)}\left(j+D\left(\rho\right)\partial_x \rho\right)^2$ and  \eqref{eq:continuity equation} is implicitly assumed. For large systems $L \gg 1$, observables  obtained from  $\mathcal{P}_t$   are governed by a saddle point approximation. This implies that obtaining the LDF amounts to solving a  minimization problem with two  constraints; the continuity equation and  particle transfer equals $Q$. Namely  $t \, \Phi \left( J \right) = L\,  \min_{\rho,j} \int dxdt' \, \mathcal{L}$, with the constraints  \eqref{eq:continuity equation}  and  $Q = L^2 \int dx d t'   \, j\left(x,t'\right)$. Moreover, we consider only density profiles with fixed boundary conditions at $x=0,1$ corresponding to  the reservoir densities. This formal minimization problem is hardly solvable in the general case as it involves solving a non-linear partial differential equation \citep{Shpielberg2016}. In \citep{Bodineau2004}, the AP conjecture was presented. It assumes that the solution to this optimization problem is time-independent, namely $j\left(x,t'\right)=J$ and $\rho\left(x,t'\right)\rightarrow \rho \left( x\right)$. As mentioned in section \ref{sec:introduction}, this conjecture is particularly  successful for boundary driven processes. The AP solution  satisfies both constraints and the LDF is now given by 
\begin{equation}
\Phi\left( J \right) = \frac{1}{L} \min_{\rho} \int ^1 _0 dx \, \mathcal{L}_J(\rho,\partial_x \rho), \label{eq:AP LDF}
\end{equation}
 with $\mathcal{L}_J (\rho,\partial_x \rho) = (J+D\partial_x \rho)^2/2\sigma$. From   \eqref{eq:AP LDF}, the LDF is recovered as a solution of an ordinary differential equation. A significant improvement to solving a partial differential equation. Throughout the text, we assume the AP solution is valid (see \citep{Shpielberg2017a} for a counter-example).  
% ======================================================================================== %
% \section{Lagrangian Mechanics Analogy
% \label{sec:Analogy}}

To present the Lagrangian mechanics analogy, let us redefine $ s=xJ$, $\tau = J$  and $W = L\Phi(J)/J$. Then,  \eqref{eq:AP LDF} becomes 
\begin{equation}
W\left(  \tau  \right) = \min_{\rho} \int ^{\tau} _0 ds \mathcal{L}_1 
(\rho,\partial_s \rho), \label{eq:action redef } 
\end{equation}
with $\mathcal{L}_1  = (1+D\partial_s \rho)^2 / 2\sigma$. $W$ can be interpreted as the minimal action of a particle to travel between the position $\rho_l$ at time $s=0$ to  the position $\rho_r$ in time $s=\tau$. Finding  $W(\tau)$  requires  solving the Euler-Lagrange equation 
%$\frac{d}{ds}\frac{\delta \mathcal{L}_1 }{\delta \partial_s \rho} = \frac{\delta \mathcal{L}_1 }{\delta \rho}$
for the trajectory $\rho(s)$ constrained at the initial and final time. 
These Dirichlet boundary conditions allow for multiple solutions. 
Assume that for $i=1,2$  there are  two solutions of the Euler-Lagrange equation denoted by $\rho_i\left(s\right)$ with  $W_i \equiv \int ds \, \mathcal{L}_1 \left(\rho_i,\partial_s \rho_i\right)$.  If there exists $\tau_C$ such that $W_1 < W_2$   for $\tau<\tau_C$ and $W_1 > W_2$   for $\tau>\tau_C$,  $W = \min{\lbrace W_1,W_2\rbrace}$ is a non-analytic function at $\tau_C$. We identify such non-analytic points as  DPTs  \footnote{Notice that in this case, W is non-convex. W must be convexified to obtain the proper solution  much like a free energy for thermodynamic phase transitions. See the full discussion in \cite{Bertini2005}.  }. 

A general solution of the Euler-Lagrange equation is hard to obtain. Moreover,  our goal is to identify DPTs  using  non-perturbative geometrical considerations,  without solving  differential equations. To do so, we consider the Hamiltonian $H$ corresponding to $\mathcal{L}_1$. With the canonical variables $\rho,\Pi$ (see appendix \ref{sec:app:Hamiltonian formalism}), we find   $H(\rho,\Pi) = E_k + V(\rho)$. Interpreting   $E_k = \frac{1}{2m}\Pi^2$  as the `kinetic energy' with the non-negative mass $m=\frac{D^2}{\sigma}$, and the `potential'  $V=-1/2\sigma$. The particle's trajectory is determined by the Hamilton equations. Let us relax the Dirichlet boundary conditions and instead use Cauchy boundary conditions to uniquely determine the trajectory $\rho\left(s\right)$. Namely, set $\rho(s=0)=\rho_l,\Pi(s=0)$ and find the corresponding $\tau$  values (if any exist) for which  $\rho\left(s=\tau\right)=\rho_r$. Varying  $\Pi\left(s=0\right)$ allows to obtain all  possible solutions. Then, we may evaluate $W$ as (see appendix \ref{sec:app:Hamiltonian formalism})  
\begin{equation}
W\left( \tau\right) = \int_0 ^\tau ds    (E_K - V +\Pi \left(-2V/m\right)^{1/2} ). \label{eq:W in canon}
\end{equation}
The Hamilton equations are by no means easier than the Euler-Lagrange equations. However, all the solutions can be identified from the geometry of the potential using the Cauchy boundary conditions.

% ======================================================================================== %

\section{Dynamical Phase transitions 
\label{sec:Dynamical Phase transitions}}

In what follows, we analyze the possible solutions of  \eqref{eq:action redef } and decide whether DPTs can occur in a toy model,  defined at the macroscopic level only. This model  highlights the advantages of the Hamiltonian approach. Physically relevant models will be discussed in section \ref{subsec:WASEP}, section \ref{sec:summary}  and in the appendix \ref{sec:app:Physical models}. The analysis goes as  follows. Identify the possible solutions for some  boundary conditions.  If more than one solution exists, evaluate $W(\tau)$ for the solutions at $\tau\rightarrow 0,\infty$. If different solutions become optimal at the different limits $\tau\rightarrow 0,\infty$, there is a DPT. 

\begin{figure}[t]
\begin{centering}
 \includegraphics[width=1\columnwidth]%{Potential_Plot.png}
{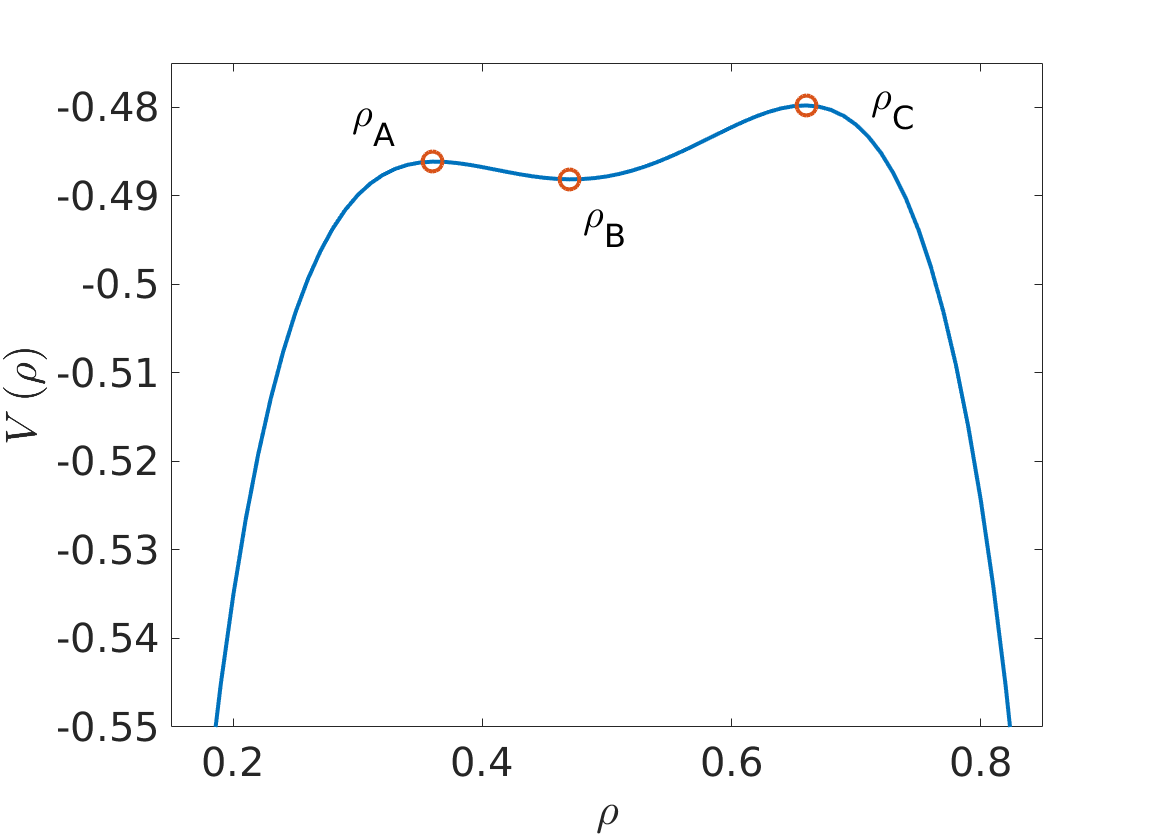}
\par\end{centering}
\caption{\label{fig:potentials} 
(Color online).
The  potential for  AMFH model with $B=-20,\epsilon=0.02$ (solid blue line). The   extremal points are marked in (red) circles, where  $\rho_A (\rho_C)$ is the lower (higher) peak and $\rho_B$ is the local minimum.}
\end{figure}

It will become clear that cyclic trajectories, i.e. $\hat{\rho}= \rho_l = \rho_r$, provide  ample intuition where to search for  DPTs. Therefore, we focus on cyclic trajectories in this Section. Two acyclic trajectories are analyzed in a similar manner in   appendix \ref{sec:app:Acylcic trajectories}.  We analyze only the cases where $\tau>0$. A similar, time reversed analysis can be made. However, due to the Gallavotti-Cohen relation \citep{Gallavotti1995}, the optimal density profiles are identical to the time-reversed solutions (for any boundary conditions). The negative current part of LDF can be inferred from the positive current part \citep{Bertini2015,Bodineau2004}. Moreover, we note that no new behavior can be obtained by switching $\rho_l \leftrightarrow \rho_r$, as this is merely conditioning the time-reversed process.  In what follows,  the sign of $\Pi$ corresponds to the direction of the $\rho$ axis (see Fig. \ref{fig:potentials}). Moreover, we identify different solutions by setting  $\Pi_0=\Pi\left(s=0\right)$ at the initial time, which, together with the initial position $\rho_l$ determines the energy $H$. In this section, only analytical mechanics arguments will be invoked. 

% ===== Fluctuation Value ===== %

The Asymmetric Mexican Flat Hat (AMFH)  model \citep{Shpielberg2017a}  is a  toy model, used here to demonstrate how to identify DPTs under the AP assumption.  We define macroscopically  $D=1$ and 
\begin{equation}
\sigma = \left(\rho-\frac{1}{2}\right)^2 + B\left(\rho-\frac{1}{2}\right)^4  -\frac{B+4}{16} + \epsilon \rho^2 (1-\rho) \label{eq:AMFH sigma}. 
\end{equation}
For $B=-20$ and $\epsilon=0.02$,  the potential of the AMFH model presents three extremal  points at  $\rho_A,\rho_B,\rho_C$ (see Fig.~\ref{fig:potentials}). We  analyze 
%the cyclic trajectories for
3 of the 7 cyclic boundary conditions cases.  The 4 remaining cases are left   to appendix \ref{sec:app:AMFH cyclic}. {A few  acyclic processes are analyzed in appendix \ref{sec:app:Acylcic trajectories} in a similar manner. 

Case 1,  $\hat{\rho} = \rho_A$:  Here, there are at least two distinct  trajectories. In the first, the particle stays put. Namely, $\rho\left(s\right)=\rho_A$ and  $\Pi\left(s\right) =0$. This solution is viable for any $\tau$. From \eqref{eq:W in canon} we find that the action associated with this trajectory is   $W_{ \mathrm{put}}\left(\tau\right)= -\tau V\left(\rho_A\right)$. In the second possible trajectory,  $\Pi_0>0$.
This  implies $\tau\left(H\right) \in \left[\tau_0,\infty \right) $  for the  energy $H\in \left( V\left(\rho_A\right) , V\left(\rho_C\right)  \right)$. Here $\tau_0>0$ is the minimal time it takes the particle to 
cross the second trajectory, going from $\rho_A$ to climb the potential hump of $\rho_C$ (never crossing it) and than travel back. %This solution implies $\tau \left(H\right)$ is non-monotonous, as $\tau \rightarrow \infty$ when $H \rightarrow V\left(\rho_A\right)^+, V\left(\rho_C\right)^-  $. This implies that for any $\tau \in \left[\tau_0,\infty\right)$ there are at least three solutions (together with the staying put trajectory). 
We can further evaluate from \eqref{eq:W in canon}  the action corresponding to the second trajectory  for $H\rightarrow V\left(\rho_c\right)$, $\Pi_0>0$ which corresponds to  $\tau \rightarrow \infty$. We find that this action is  $W_2 \left(\tau\right)= -\tau V\left(\rho_C\right) + \mathcal{O}\left(1\right)$ 
as the particle spends most of the time approaching $\rho_C$ with vanishing $\Pi$ (see Fig.~\ref{fig:trajectories}).  
While for short times $W_{ \mathrm{put}}$ is the sole and hence  the dominant trajectory, we found that  for some large but finite $\tau$,  the second trajectory dominates. Therefore $W$ is non-analytic and a DPT takes place. 
We note that such a DPT could not be obtained using a perturbative approach  \citep{Baek2016b,Shpielberg2016}.  

Case 2, $\hat{\rho} < \rho_A$:
Here there are again at least two possible trajectories, both with $\Pi_0>0$. We denote the first trajectory as the short path. It corresponds to $\tau_{ \mathrm{short}}\left(H\right) \in \left[0,\infty \right) $  for $H\in \left[ V\left(\hat{\rho}\right) , V\left(\rho_A\right)  \right)$, where the particle never crosses the lower potential peak at $\rho_A$.  In the long path, the particle crosses the potential peak at $\rho_A$, but not the one at $\rho_C$. This corresponds to $\tau_{ \mathrm{long}}\left(H\right) \in \left[\tau_0,\infty \right) $  for $H\in \left( V\left(\rho_A\right) , V\left(\rho_C\right)  \right)$. $\tau_0$ is the minimal  finite time it takes the particle to complete the long trajectory (not the same value as in case 1).    At long times, we can evaluate the action of the short path using \eqref{eq:W in canon}. We find $W_{ \mathrm{short}} = -\tau V \left(\rho_A\right) + \mathcal{O}\left(1\right)$. Similarly, we can evaluate the action of the long path for $H\rightarrow V\left(\rho_C\right) $. We find $W_{ \mathrm{long}} = -\tau V \left(\rho_C\right) + \mathcal{O}\left(1\right)$. So,  while $W_{ \mathrm{short}}$ dominates at short times $\tau<\tau_0$, $W_{ \mathrm{long}}$ dominates at long times and a DPT must occur.  

Case 3, $\hat{\rho}=\rho_B$:  
Here, there are infinitely many distinct trajectories. Let us characterize them. 
The first possibility is staying put as the particle sits on a potential extremum. The solution is valid for any time $\tau$ with an associated action $W_{ \mathrm{put}}\left(\tau\right)= -\tau V\left(\rho_B\right)$. We note that this is the only trajectory allowing for $\tau \rightarrow 0$. A second possible solution is for the particle to pick up some initial positive  (negative) momentum $\Pi_0$. Even for infinitely small momentum, the minimal time $\tau_0$ to cross the path (going slightly up the potential and back down to $\rho_B$) is finite. The motion is essentially that of half a  cycle of an harmonic oscillator. Solving the harmonic oscillator equations of motion around the point $\rho_B$, we find that $\tau_0 = \pi \sqrt{\frac{m(\rho_B)}{\partial_{\rho \rho}V(\rho_B)}}$. If a transition between the 'staying put' trajectory to this one occurs, it can be a  continuous transition. A transition indeed occurs in this case exactly at $\tau_0$.  it  can be validated by numerically   solving the Hamilton equations (Appendix \ref{sec:app:numerics}), or analytically using a perturbative approach \citep{Baek2016b}. We note that one can  evaluate the order of magnitude of   $\tau_0$ by using a harmonic oscillator approximation for any of the cases studies for the AMFH model. We can also consider multiple crossings of $\rho_B$ as long as the energy does not exceed $V\left(\rho_A\right)$. 
Considering $\Pi_0>0$ with $H\rightarrow V(\rho_C)$ amounts to $W_\mathrm{right}\left(\tau \right) =  - \tau V(\rho_C) +\mathcal{O}\left(1\right) $ for $\tau \rightarrow 0$. This trajectory is certainly preferable to the 'staying put' trajectory and thus we may have a richer phase diagram. While the geometrical approach does not help to obtain  the full phase diagram, it certainly verifies that  a DPT indeed occurs in this scenario.

A numerical verification for the DPTs found here and in  appendix \ref{sec:app:AMFH cyclic} was performed and is discussed in appendix \ref{sec:app:numerics}. We also note that as in other cases, multiple transitions may occur. Case 2 dispels any illusion that DPTs are related to boundary conditions on or close to extremal points of $\sigma$. 

To conclude, we found that extremal points in the potential (conductivity) are  facilitators of DPTs. It should be clear that similar arguments can be invoked to identify DPTs for acyclic trajectories. In the corresponding MFT picture, the particle trajectories are density profiles. Fig.~\ref{fig:trajectories} depicts the density profiles  for small and large positive values  of the  current $J$ (short and long time $\tau$) for the first two cases discussed in this Section.

% ===== DPT trajectories ===== %
\begin{figure}[t]
\begin{centering}
 \includegraphics[width=1\columnwidth]{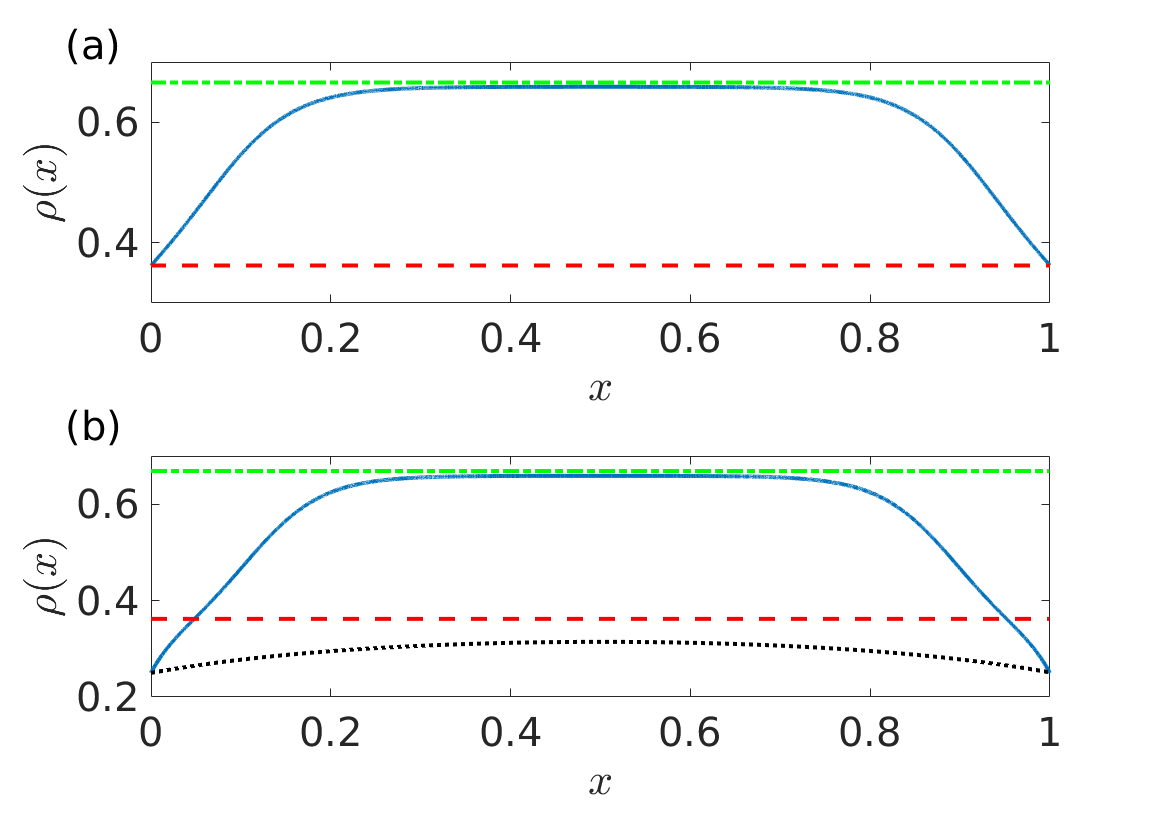}
 \par\end{centering}
\caption{\label{fig:trajectories} 
(Color online).
Typical density profiles in the AMFH model. (a) Case 1, $\hat{\rho}=\rho_A$, the (red) dashed line corresponds to the staying put trajectory and the (blue) solid line corresponds to the 2nd trajectory at large currents (large $\tau$). It begins to saturate $\rho_C$ represented by the (green) dashed-dotted line as expected. (b) Case 2, $\hat{\rho}<\rho_A$, the (blue) dotted line corresponds to the short trajectory for low values of the current ( $\tau<\tau_0$) and the (red) solid line corresponds to the long trajectory at large currents ($\tau>\tau_0 $). It begins to saturate $\rho_C$ represented by the (green) dashed-dotted as expected.      }
\end{figure}

% ======================================================================================== %

% ======================================================================================== %

\section{Generalization to models with weak fields
\label{sec:weak fields}}

Let us now consider the macroscopic fluctuation theory for models with a weak asymmetry in the form of a field of strength $E$. The Einstein relation implies that Fick's law \eqref{eq:Fick} is modified to include the asymmetry by the addition of a  $E\sigma$ term. Repeating the same procedure of the fluctuating hydrodynamics and assuming the AP, we find  that the Lagrangian in \eqref{eq:AP LDF} is modified to $\mathcal{L}_J = (J+D\partial_s \rho + E \sigma)^2 /2\sigma$. Then, going to the Lagrangian mechanics, we find that \eqref{eq:action redef } is given with $\mathcal{L}_1 = (1 + D\partial_s \rho + \tilde{E}\sigma)^2 /2\sigma$ with the rescaled field $\tilde{E} = E/\tau $. Thus, in principle, the potential changes as a function of the end-time $\tau$. This in turn allows to identify more DPTs, albeit in a more subtle way. Moreover, notice that for $\tau \rightarrow \infty $, the action loses its dependency of the field $E$. The Hamiltonian here is given by the same form with the same mass term. The generalized momentum $\Pi =  \frac{\partial \mathcal{L}_1}{\partial \partial_s \rho} - \frac{D}{\sigma}- D \tilde{E}$ and the potential is $V\left(\rho\right)=- \frac{1}{2\sigma}\left(1+\tilde{E}\sigma\right)^2$. The action $W$ in \eqref{eq:W in canon} remains of the same form. Our course of action will be only slightly different than the zero field case. First, we will identify initial and final conditions where for $\tau \rightarrow  \infty$  and $\tilde{E}\rightarrow 0$ there is a single solution. Then, for finite $\tau$ and large $\tilde{E}$, we  identify another solution in addition to the large $\tau$ solution. This new solution will be argued to dominate, namely it has a smaller action. As before, this guarantees a DPT. It should be noted that while there is freedom in the selection of $E$, once chosen, it is kept fixed throughout the process, at least in our specific setup. 
% ======================================================================================== %

  \subsection{The Weakly Asymmetric Simple Exclusion Process
  \label{subsec:WASEP}}
 
  The Simple Symmetric Exclusion Process (SSEP) is a paradigm process for non-equilibrium systems as it is solvable by Bethe ansatz \citep{Derrida1993,Derrida2007}. In the SSEP, there is at most one particle per  site, and particles hop to empty  neighbors with rate $1$. This implies $D=1$ and $\sigma=2\rho\left(1-\rho\right)$. The large deviation function is known to be analytic, so no DPTs occur.   In the Weakly Asymmetric Simple Exclusion Process (WASEP), particles hop to empty sites to the right (left) with rate $1\pm E/L$. The scaling of the field with the system size keeps the process diffusive. To identify DPT, we draw the potential of the WASEP for several values of $\tilde{E}$ (see Fig.~\ref{fig:WASEP_pot}). We specify two cases of interest.

Case 1, $\hat{\rho} = 1/2$: In the SSEP, there is an obvious particle hole symmetry and therefore this is an immediate point of interest. For $\tilde{E}\rightarrow 0$ which corresponds to the limit $\tau \rightarrow \infty$, there is a single trajectory. Namely, staying put $\rho(s)= 1/2$ with $\Pi(s)=0$ (see Fig.~\eqref{fig:WASEP_pot}). For $\tilde{E}\gg 1$ there are more solutions. Staying put remains a solution. Another solution is for the particle to start climbing the potential and then tumble back. We have already seen in Section \ref{sec:Dynamical Phase transitions} case 3, that if $\tau$ is large enough this solution is dominant. Since we still have the freedom to choose $E$, we can always have  $\tilde{E}$ large  while keeping a large enough $\tau$. Thus, we expect a DPT at large values of $E$. This case was also discussed in \citep{Baek2016b}.

Case 2, $0 <\rho_l < 1/2$ and $\rho_r = 1-\rho_l > 1/2$: For $\tilde{E}\rightarrow 0$ which corresponds to the limit $\tau \rightarrow \infty$, there is again a single trajectory. The particle must have $\Pi_0>0$ large enough to overcome the potential barrier. However, for  $\lvert \tilde{E}\rvert > 2 $, the potential is completely changed as two new maxima appear (at $\rho_{\pm} = \frac{1\pm \sqrt {1-2/\lvert \tilde{E}\rvert}}{2} $) and the old maximum becomes a minimum. Now, suppose that $\tilde{E}$ is such that $\rho_l,\rho_r$ are located between the two maxima. Then, we may focus on two trajectories. The direct one with $\Pi_0>0$ and the indirect where the particle has  $\Pi_0<0$ and it starts climbing towards the left maximum of the potential. For the direct path, $W_\mathrm{direct} \geq  -\tau V(\rho_l)$, as $E_k,\Pi>0$ in this trajectory. For the indirect path at the long time limit, we can evaluate $W_\mathrm{indirect}=-\tau V((\rho_{\pm}))+ \mathcal{O}(1)$. Since we have picked $\rho_l$ such that $V((\rho_{\pm}))>V((\rho_{l}))$ a DPT must occur as $W_\mathrm{indirect}<W_\mathrm{direct}$. So, like in case 1 above, we can always choose $E$  that allows $\tau$ to be large enough for the desired value of $\tilde{E}$. This guarantees a DPT. In fact, we need not require any symmetry between $\rho_l$ and $\rho_r$ to observe the transition as we choose $\tilde{E}$ such that  $V((\rho_{\pm}))>V((\rho_{l,r}))$. Therefore, a transition should be expected for any choice of $\rho_l<\frac{1}{2}<\rho_r$ for a sufficiently large $E$.  

The  large deviation function of the SSEP is known to be analytic \citep{Derrida2004,Bodineau2004,Imparato2009}. We have seen that applying a weak field may generate new extremal points in the effective potential, thus allowing for DPTs. However,  adding a weak field to a model  will not necessarily  generate DPTs for any model. One such counter example is for non-interacting Random walkers, where a single solution was obtained for any weak field \citep{Hirschberg2015}. While the potential may posses a local maximum point, it is not enough to identify a DPT using the geometrical approach. 

% ===== potential for KLS and Bodineau ===== %
\begin{figure}[t]
\begin{centering}
 \includegraphics[width=1\columnwidth]{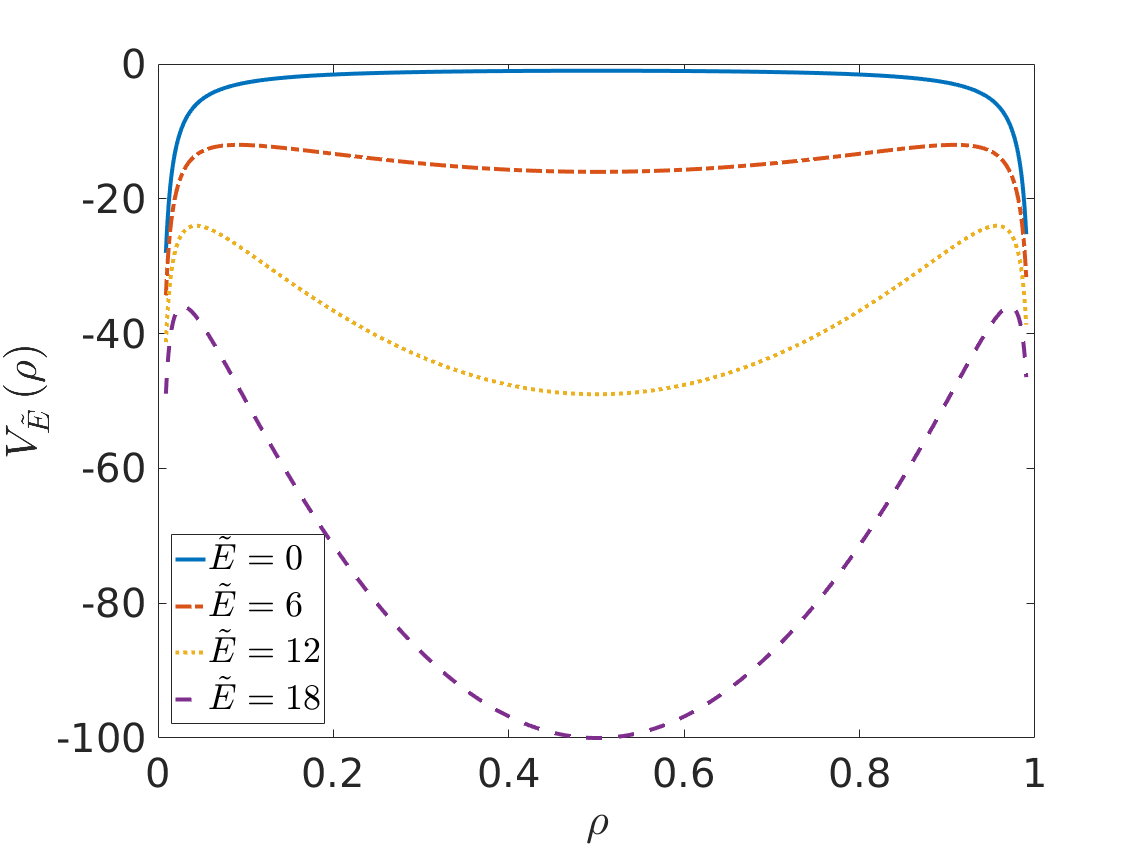}
\par\end{centering}
\caption{\label{fig:WASEP_pot} 
(Color online).
The corresponding potentials for various values of $\tilde{E}$ of the WASEP. It can be seen that the basic structure of the potential is completely changed for different values of $\tilde{E}$. For   $\lvert\tilde{E}\rvert > 2$, there are two maxima at $\rho_{\pm}=\frac{1\pm \sqrt{1-2/\lvert \tilde{E}\rvert}}{2}$ and one minimum at $\rho=1/2$, whereas for $|\tilde{E}|< 2$ there is a single maximum at $\rho=1/2$. This behavior give rise to the DPTs discussed in the main text. }
\end{figure}

%   with $V\left(\rho\right) = -\frac{1}{4\rho\left(1-\rho\right)}$ shown in Fig.~\ref{fig:potentials}. 
 
% Here the potential has  a maximum, at $\rho=1/2$.

% ======================================================================================== %

\section{Discussion
\label{sec:summary}}

We have presented here a mapping between current fluctuations in boundary driven systems under the AP assumption to the evolution of a Hamiltonian particle with set initial and final  positions.  We have then shown that a pictorial analysis of the potential is sufficient to demonstrate DPTs. Note that the Hamiltonian  approach allows to focus on the geometry of the potential, rather than exact details. While the AMFH model is  a toy model, the conclusions presented above apply for  models with similar conductivity, e.g. Bodineau's long-range hopping model \citep{Shpielberg2016,Shpielberg2017a} and the Katz-Lebowitz-Spohn model \citep{Katz1983,Katz1984,Hager2001,Baek2016b}. See appendix \ref{sec:app:Physical models} for details on these two processes.  
This approach allows to obtain new DPTs, as well as derive on simple terms known DPTs \citep{Baek2016b,Shpielberg2017a}. 

The geometrical approach provides a sufficient condition for DPTs. We found that extremal points in the potential are   facilitators of DPTs. One can wrongly assert that, since the diffusion does not play a role in the potential, it is immaterial to the study of DPTs.  
%The diffusion plays no role if $\tau \rightarrow \infty$ with bounded $H$ (due to an extremal point in the potential). 
However, aside from affecting the value of the critical current (or corresponding time $\tau$), the diffusion may allow for a richer phase diagram (see cases D,E in appendix \ref{sec:app:AMFH cyclic}  and \citep{Baek2016b} for examples). It cannot be ruled out that the non-trivial mass term facilitates a DPT that cannot be identified from the potential. Such DPTs are outside the scope of this paper and are not neatly described by the formalism presented here. 
%\textcolor{blue}{ A rich class of transitions can be found in case the mass term rapidly changes (especially if the potential does not). A heavy particle has a lower action. Thus, it may be interesting to build toy models supporting such a scenario. More interesting cases are for non-analytic $D,\sigma$ themselves (see \citep{Krapivsky2013a}).  }

We further  note that once a DPT is identified using the Lagrangian approach, the order of magnitude of the transition ($\sim \tau_0$) can be recovered from  dimensional analysis of the mass and potential (Section \ref{sec:Dynamical Phase transitions}, Case 3). 

The method was extended   to include weak driving fields. We have shown that the potential explicitly depends  on the ratio of the applied field and the current. This allows for more extremal points to be generated for different values of the field and thus it enables more DPTs. 

In appendix \ref{sec:Two species}, we have generalized the geometrical approach to the case of $d$ different species of particles. This correspond to Lagrangian mechanics of a single particle in $d$ spatial dimensions. Similarly to the case of  a boundary driven process with a weak field, more control parameters are included (more than one constrained current). This implies that in principle, DPTs should be found in abundance for physical models with interacting particle species as the control parameters can be used to generate more extremal points in the potential. 
% \textcolor{blue}{Moreover, DPTs can be found also for models outside the scope of the MFT. A hydrodynamic description was given for a run and tumble process of bacterial dynamics \citep{Thompson2011}. While that specific model may not be sufficient to exhibit a DPT for current fluctuations, a model in the same spirit including exclusion of particles and a large parameter to facilitate the saddle point, is sufficient to identify a DPT.     }

Finally, an experimental setup realizing the analog of the LDF can be considered.  Direct experimental measurement of the LDF is hard as  we are searching for exponentially rare events. Finding a  mechanical system  described by the effective Hamiltonian $H$ allows to  experimentally probe $W(\tau)$, the equivalent of the LDF. As the mapping suggests atypical currents $J\rightarrow \tau$, we find that an analog Hamiltonian explores large deviations in linear time. This exponential reduction is due to the AP, allowing to discard many trajectories (see \citep{Nemoto2014} for similar motivation). 
%One is required to monitor the potential energy of the trajectory as  a function of the initial total energy $H$. 

One possible realization is by lacing a bead of mass $m$, susceptible to gravity $g$, through a hard string. Negligible dissipation of energy is assumed throughout the process. The string's contour is given by $\vec{r}=(x,f(x),h(x))$. Thus, the corresponding Hamiltonian is  $H= \frac{1}{2m} (\Pi_x ^2 +\Pi_z ^2 +\Pi_z^2	) - mgz   $. Hamilton equations dictate  $\Pi_y =  \Pi_x \partial_x f(x) $ and $\Pi_z =  \Pi_x \partial_x h(x) $. This amounts to rewriting an effective $1d$  Hamiltonian $H_{ \mathrm{eff}} = \frac{1}{2m_{ \mathrm{eff}}} \Pi_x ^2 - mgh(x)   $, with $m / m_{ \mathrm{eff}}  = 1+\left(\partial_x f\right)^2 +\left(\partial_x h\right)^2$. Control over $f,h$ allows to replicate the desired space-dependent effective mass and potential in a finite range  for a variety of $D,\sigma$ functions (see appendix \ref{sec:app:Experimental modeling}).

Note that there is no clear advantage to finding $W$  experimentally rather than a numerical evaluation. However, this example  shows that DPTs could be observed experimentally in linear time using analog systems. This idea motivates searching for the equivalent of the AP in other large deviation problems. 

%if the mechanism of the transition is understood.      

% ======================================================================================== %

\begin{acknowledgments}
This work has been supported by ANR-14-CE25-0003. OS would like to thank Yaroslav Don, Denis Bernard, Takahiro Nemoto, Boris Timchenko and Jan Troost   for useful discussions comments and ideas. 
\end{acknowledgments}

% ======================================================================================== %

% \clearpage

% a small hack to start the appendix leveled
\newpage
\begin{widetext}
\end{widetext}

% === Appendix === %

\appendix

% ======================================================================================== %

\section{Hamiltonian formalism
\label{sec:app:Hamiltonian formalism}}

In this section we derive, for completeness, the Hamiltonian $H$ corresponding to the Lagrangian $\mathcal{L}_1$ 
of the main text. we define, as usual, $H$ as the Legendre transform of $\mathcal{L}_1$. Namely, $ H = p\partial_s  q - \mathcal{L}_1 $, with $ p = \frac{\partial \mathcal{L}_1}{\partial \left( \partial_s \rho\right)}$. We find 
 \begin{equation}
 H = \frac{1}{2m} \Pi^2 + V,
 \end{equation}
where $\Pi = p - D/\sigma$, $m=D^2/\sigma$ and $V = -1/2\sigma$. Notice $\rho,\Pi$ are canonical. Defining $E_k =  \frac{1}{2m}\Pi^2 $  as the  kinetic    term  allows to identify (as usual) the total energy as the sum of the kinetic and potential energies. Note that $\Pi = m \partial_s \rho  $, which implies that zero kinetic energy makes for vanishing `velocity' $\partial_s \rho$. We can thus rewrite the Lagrangian (in an unusual way) 
\begin{equation}
\mathcal{L}_1 = E_k - V + \Pi \sqrt{-2V/m}.
\end{equation}

\section{Analysis of equal boundary conditions for the AMFH model
\label{sec:app:AMFH cyclic}}

This section deals with the case of equal boundary condition for the AMFH model, \ie $\hat{\rho} = \rho_l = \rho_r$. In the main text, three out of possible seven cases were discussed. Here, we complete the discussion by analyzing the remaining four cases. $\rho_A,\rho_B,\rho_C$ are depicted in   Fig.~\ref{fig:potentials}  of the main text.

Case 4, $\rho_A<\hat{\rho}<\rho_B$: 
Similarly to Case 3, we find in Case 4 infinitely many   solutions as the particle may revisit $\hat{\rho}$ several times. Let us focus on two solutions: the short-left and short-right. In the short-left solution, $\Pi_0\leq 0$. $\tau_{ \mathrm{left}}\left(H\right)\in \left[0,\infty  \right)$ for $H\in \left[ V\left(\hat{\rho}\right)  , V\left(\rho_A\right) \right)$.  In the short-right solution,    $\Pi_0>0$ and  $\tau_{ \mathrm{right}}\left(H\right)\in \left(\tau_0,\infty  \right)$ for  $H\in \left( V\left(\hat{\rho}\right)  , V\left(\rho_C\right) \right)$. $\tau_0>0$ is the minimal  finite time for the particle to traverse the short-right trajectory. For $\tau \rightarrow \infty$, the particle traveling in the short-left (right) trajectory spends most of its time approaching the peak at $\rho_A (\rho_C) $ with vanishing kinetic energy. In the similar manner to the main text, evaluating \eqref{eq:W in canon} implies  $W_{ \mathrm{left}}>W_{ \mathrm{right}}$ as $\tau \rightarrow \infty$.

One can also consider trajectories that cross $\hat{\rho}$ more than once as the particle can perform an oscillatory motion.  They compose an  infinite set of solutions. To find the complete phase diagram, one has to pursue a detailed analysis. However, only the short-left trajectory is viable for $\tau\rightarrow 0$. Since the short-right solution dominates over the short-left solution at long times, a  DPT is guaranteed. 

Note that here, the  phase diagram may be richer due to the infinite set of solutions at intermediate times. A detailed analysis to recover the full phase diagram is not  attempted here.

Case 5, $\rho_B<\hat{\rho}<\rho_C$: 
This case is very similar to Case C.  Here however, there is no guarantee for a DPT, as for short and long times the short-right path is favorable ($\Pi_0>0$). The intermediate times must be analyzed with care and cannot be inferred from this simple picture.

Case 6, $\hat{\rho}=\rho_C$: 
Here there is only one possible solution, staying put. The particle never returns to $\rho_C$ for any non-zero  $\Pi_0$. So, no DPT is expected.  

Case 7, $\hat{\rho}>\rho_C$:  
Here again there is only  a single possible solution with $\Pi_0<0$. Therefore, no DPT can be identified from the potential alone.

\section{Experimental modeling
\label{sec:app:Experimental modeling}}

We have shown in the main text that  hard string laced through a bead can gives a prescription for a  desired $1d$ effective dynamics.  The purpose is to find for arbitrary  $D,\sigma$, the contour of the string giving rise to the  effective Hamiltonian.  First, notice that the important parameters in the experiment are the mass of the bead $m$ the gravity constant $g$ and a characteristic length scale $x_0$. Therefore, we  attribute dimensions to $D(x)$ and $\sigma(x)$  for the model to make physical sense, 
\begin{subequations}
\begin{align*}
\sigma(x) = \frac{1}{mgx_0} \sigma(\rho) \\
D(x) = \frac{1}{\sqrt{gx_0}} D(\rho),
\end{align*}
\end{subequations}
with $\rho = x/x_0$,  $D(\rho)$, and $\sigma(\rho)$  dimensionless parameter and functions. We can thus write $h(x) = x_0 h(\rho)$ and $f(x) = x_0 f(\rho)$
Since $V = mgh(x) = 1/2\sigma(x)$ we find $h(\rho)=1/2\sigma(\rho)$.  We use the effective mass equation to find  $\partial_\rho f(\rho)$ by 
\begin{equation}
\left(\partial_\rho f(\rho)\right)^2 = \frac{\sigma(\rho)}{D^2 (\rho)} - 1 -\frac{1}{4}\frac{\left(\partial_\rho \sigma (\rho)
\right)^2}{\left(\sigma(\rho)\right)^4}. \label{eq:f contour}
\end{equation}

Unfortunately, the right hand side of \eqref{eq:f contour} is not always positive. For example, for the AMFH model, we find that the right hand side of \eqref{eq:f contour} is in fact always negative. However, recall that the DPTs are dominated by the potential, and the role of the diffusion $D(\rho)$ is secondary. So, changing $D$ to e.g. $D= \left[\left(1-\rho\right) \rho \right] ^4$ allows to find a real function $f(\rho)$ in for any $\rho \in \left[0,1\right]$.  We note that the potential can never be truly mimicked as $V(\rho \rightarrow 0) \rightarrow -\infty$ is  experimentally unreachable. This Toy model  provides merely a  proof of principle. One can compose a variety of potentials using e.g. electric fields to try and mimic the desired Hamiltonian for arbitrary $D,\sigma$. This will not be attempted here.

% ======================================================================================== %

\section{Acylcic trajectories for the AMFH model 
\label{sec:app:Acylcic trajectories}}

To complete the discussion in \ref{sec:Dynamical Phase transitions} we discuss possible DPTs for two cases of acyclic paths in the AMFH model. Namely, the reservoirs are taken at different densities. 

Case 1, $\rho_l < \rho_A $ and $\rho_A < \rho_r < \rho_B $: Here, there is always a direct trajectory where $\Pi_0>0$ with $H\in \left( V\left(\rho_A\right), \infty \right)$. The particle passes over the $\rho_A$ potential peak and directly continues to $\rho_r$. Here, larger energies $H$ correspond to smaller time values $\tau $. Any $\tau$ value is viable. A second possible solution is again for $\Pi_0>0$, with $H\in \left(V\left(\rho_A\right),V\left(\rho_C\right)\right)$. Here the particle crosses $\rho_l$ once as it starts to climb towards $\rho_C$, only to tumble back down towards $\rho_l$. Here there is some minimal finite time $\tau_0$, below which the trajectory cannot be realized. This ensures that the direct  trajectory dominates at short times. Evaluating \eqref{eq:W in canon} ensures that the direct trajectory is no longer dominant at large enough times. Therefore, we have identified a DPT. 

Case 2, $\rho_l = \rho_A$ and $\rho_r = \rho_C$: Here it is easy to understand that there is only a single trajectory possible to reach from $ \rho_A$ to  $\rho_C$. This means we cannot identify a DPT from the geometrical approach. Of course, this does not exclude a DPT altogether.

% ======================================================================================== %

\section{Physical models that support DPTs
\label{sec:app:Physical models}}

Let us present here in more details two physically relevant models, that reproduce the discussed DPTs of the main text. 
\subsubsection{The long-range hopping with exclusion model}
% \begin{description}
%   \item[The long-range hopping with exclusion model] \hfill \\ 
  This model,  proposed by
Bodineau \citep{Shpielberg2017a,Shpielberg2016}, is a one-dimensional lattice-gas model with at most one particle per lattice site. A particle
can hop from site $i$ to an empty nearest-neighbor site $i \pm 1$ with rate
$1$ and it is also allowed  to hop from site $i$ to an empty site $i \pm (\beta + 1)$ with  rate
$\alpha$ provided that the $\beta$ sites separating them are all occupied.  $D$ and $\sigma$ can be obtained analytically as this this is a gradient model \citep{Spohn1992Part2Chap2}. We obtain $D(\rho) = 1 +
\alpha (\beta + 1)^2  \rho^\beta $ and $\sigma (\rho) = 2\rho (1 -\rho ) D(\rho)$ with  $\rho \in [0,1]$. Choosing $\alpha = \frac{1}{24}$ and $\beta = 9 $ allows to reproduce the double peaked potential as shown in Fig. \ref{fig:Bod_KLS_potentials}. This of course allows to reproduce the DPTs discussed in the main text.

\subsubsection{The Katz-Leibowitz-Spohn model }
  
%   \item[The Katz-Leibowitz-Spohn model ] \hfill \\ 
The Katz-Leibowitz-Spohn \citep{Katz1984,Katz1983} model is a lattice gas model with exclusion, that incorporates nearest-neighbor hopping with interactions. 
The dynamics of right-handed hopping is given below, where full circles represent occupied sites and empty circles represent empty sites. 
\begin{eqnarray}
\circ \, \bullet  \, \circ \, \circ  \xrightarrow{{1+\delta}} \circ \, \circ  \, \bullet \, \circ   
% 0100 ⟶ 0010;  1+δ
&  \qquad  &
\bullet \, \bullet  \, \circ \, \circ  \xrightarrow{{1+\epsilon}} \bullet \, \circ  \, \bullet \, \circ 
% 1100 ⟶ 1010;% 1+ε
\nonumber \\
\bullet \, \bullet  \, \circ \, \bullet  \xrightarrow{{1-\delta}} \bullet \, \circ  \, \bullet \, \bullet  
% 1101 ⟶ 1011; 1−δ
&  \qquad   &
\bullet \, \circ  \, \bullet \, \circ  \xrightarrow{{1-\epsilon}} \circ \, \bullet  \, \bullet \, \circ 
% 1010 ⟶ 0110: 1−ε
\nonumber
\end{eqnarray}
The spatially inverted transitions occur with identical rates. The parameters $\epsilon,\delta$ provide some control over $D,\sigma$, where exact expressions  are given in the appendix of \citep{Baek2016b}. By choosing, e.g. $\epsilon=0.99,\delta=0.45$, we can obtain the desired double peaked potential to reconstruct the DPTs discussed in the main text (see Fig.~\ref{fig:Bod_KLS_potentials})
% \end{description}
%epsilon = 0.99 ; delta = 0.45 ; 

% put the figure 
% ===== potential for KLS and Bodineau ===== %
\begin{figure}[t]
\begin{centering}
 \includegraphics[width=1\columnwidth]{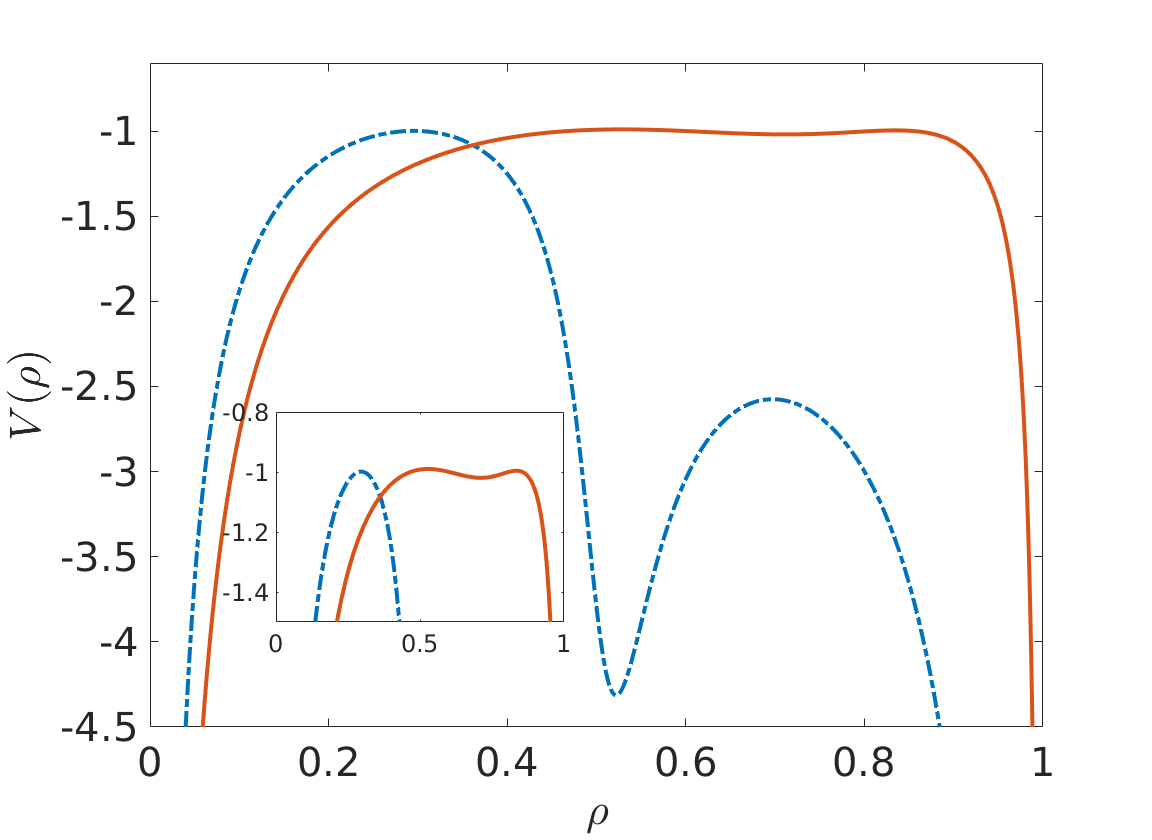}
\par\end{centering}
\caption{\label{fig:Bod_KLS_potentials} 
(Color online).
The corresponding potentials for the Long-range hopping model with $\alpha  = \frac{1}{24}, \beta = 9$ (solid 
red line) and the KLS model  with $\delta =0.45,\epsilon=0.99$ (dashed  blue line). The inset  shows a zoom-in on the extremal points structure of the Long-range hopping model.}
\end{figure}

% ======================================================================================== %

\section{Multi-species models 
\label{sec:Two species}}
In this section we show how to extend the Hamiltonian approach to one dimensional boundary driven systems with $d$ different species of particles. Here again, we consider all the different species are conserved in the bulk, namely $\partial_t \rho_\alpha = -\partial_x j_\alpha$ for $\alpha = 1...d$. We also consider a generalized stochastic Fick's law of the form $j_\alpha = q_\alpha + \sqrt{\frac{\sigma_{\alpha \beta}}{L}} \xi_\beta$, where  the conductivity $\sigma$ determines the strength of the fluctuations and $q_\alpha$ determines the mean current. To make matters simple, we consider  $q_\alpha$ of the form 
\begin{equation}
q_\alpha =  - D_{\alpha \beta} \partial_x \rho_\beta,
\end{equation}
where $D_{\alpha \beta} $ is the diffusion matrix. We can thus obtain a Lagrangian of the form 
\begin{equation}
\mathcal{L}= \frac{1}{2}\sigma^{-1} _{\alpha \beta} (j_\alpha+q_\alpha)(j_\beta+q_\beta),
\end{equation}
where $\sigma^{-1}$ is the inverse of the conductivity matrix. 

Now, using the AP, we take $j_\alpha \left(x,t\right) \rightarrow J_\alpha$ and $\rho_\alpha \left(x,t\right)\rightarrow \rho_\alpha \left(x\right)$. As usual, we rescale $x\rightarrow s = xJ_1\in \left[ 0,\tau\right]$, such that we want to find 
\begin{equation}
W\left(\tau\right) = \int ^\tau _0 ds \, \mathcal{L}_{1d},
\end{equation}
where $\mathcal{L}_{1d} = \frac{1}{2}\sigma_{\alpha \beta} \left(r_\beta +q_\beta \right) \left(r_\beta +q_\beta \right)$, with $r_\alpha = J_\alpha / J_1$. The corresponding Hamiltonian to $\mathcal{L}_{1d}$ is 
\begin{equation}
H_{1d} = E_K + V,
\end{equation}
with 
\begin{eqnarray}
E_K &=& \frac{1}{2}m^{-1} _{\alpha \beta} \Pi_\alpha \Pi_\beta \\ \nonumber
V& =&  - \frac{ 1}{2} r_\alpha  r_\beta    \sigma^{-1} _{\alpha \beta}.
\end{eqnarray}
Where we have defined 
\begin{equation}
m_{\mu \nu} =  D_{\alpha \mu } \sigma^{-1} _{\alpha \beta} D_{\beta \nu }.
\end{equation}
As usual, $\Pi_\alpha$ are canonical variables to $\rho_\alpha$  and $W\left(\tau\right)$ is also given by 
\begin{equation}
\mathcal{L}_{1d} = E_k - V + \sigma^{-1} _{\alpha \beta} D_{\alpha \gamma } m^{-1} _{\eta \gamma} r_\beta \Pi_\eta. \label{eq:spec Action W} 
\end{equation}
One can verify that \eqref{eq:spec Action W} becomes \eqref{eq:W in canon} for a single species of particles. Identifying a DPT can be done in a similar fashion to what was done in sections \ref{sec:Dynamical Phase transitions},\ref{sec:weak fields}. However, finding and analyzing a microscopic model that presents such a transition is beyond the scope of this paper.

% ======================================================================================== %

\section{Numerical verifications 
\label{sec:app:numerics}}

In this section we present  numerical verifications for cases 1,3 of the AMFH model  in Section \ref{sec:Dynamical Phase transitions}. We numerically solve the Hamilton equations with the Cauchy boundary conditions. Namely, we set the initial conditions $\rho(s=0)= \rho_l$ and $\Pi(s=0)=\Pi_0$. We vary $\Pi_0$ to set the energy $H$ within the allowed range of the desired trajectory. We identify $\tau$ to satisfy $\rho(\tau)=\rho_r$. Note that for a trajectory of choice we already determine how many times $\rho(s)$ visits $\rho_r$ as $\tau$ need not be the first time $\rho(s)=\rho_r$.

%We explicitly solve the Hamilton equations according to the identified trajectories, with respect to the allowed energy range. 
The plots of $W(\tau)$ for the different trajectories are shown in Fig.~\ref{fig:W cases 1 3}.  
% put the figure 
% ===== W values for case 1 ===== %
\begin{figure}[t]
\begin{centering}
 \includegraphics[width=1\columnwidth]{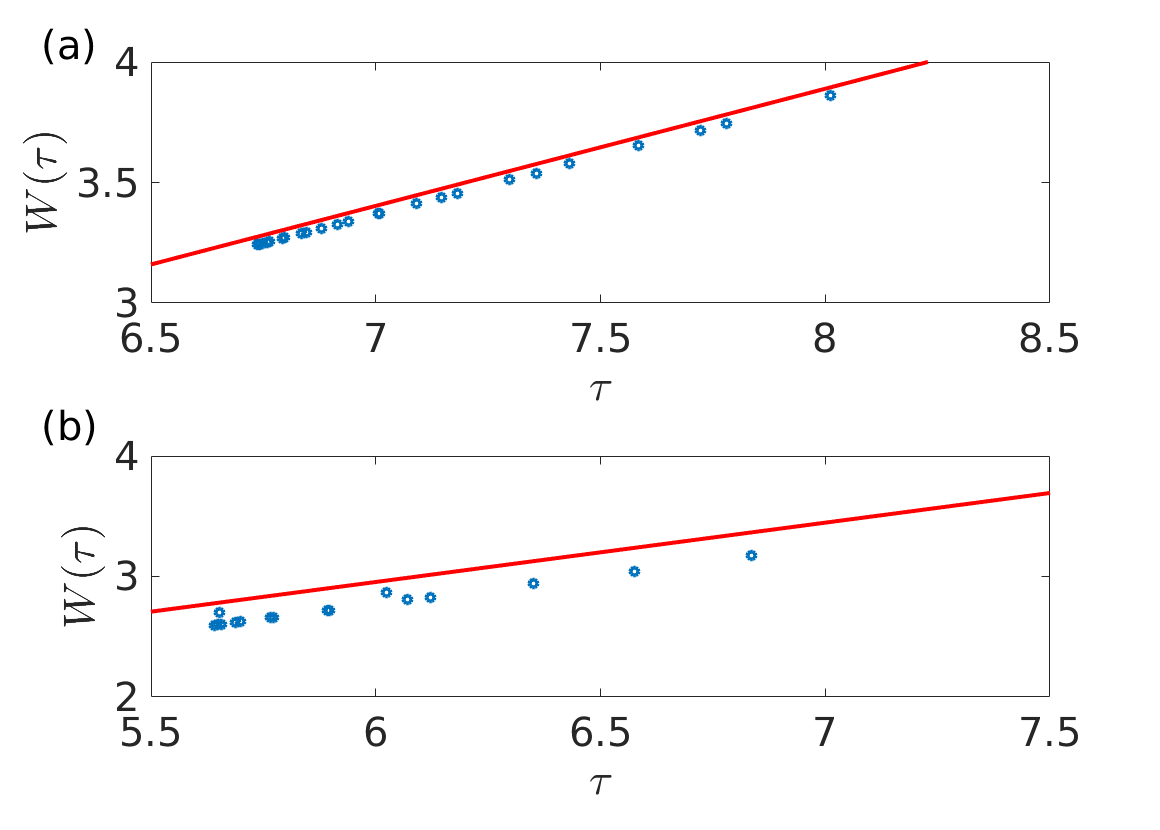}
\par\end{centering}
\caption{\label{fig:W cases 1 3} 
(Color online).
The numerical values for the action $W(\tau)$ for two possible solutions are considered. (a) corresponds to Case 1 and (b) to Case 3 in Section \ref{sec:Dynamical Phase transitions}. 
The (red) curve represents the action $W(\tau)$ in the `staying put' solution and the (blue) circles correspond to the action $W(\tau)$ of the positive initial momentum $\Pi_0$ solution. In both cases, the positive initial momentum solution becomes dominant as soon as it is feasible (i.e.  $\tau_0=6.7 (5.6)$ for case 1 (3) correspondingly). }
\end{figure}

% % ===== W values for case 2 ===== %
% \begin{figure}[t]
% \begin{centering}
%  \includegraphics[width=1\columnwidth]{W_below_VA.png}
% \par\end{centering}
% \caption{\label{fig:W case 2} 
% (Color online).
% The (red) circles correspond to the action of the short-path and the (blue) dots to the action of the long-path. In the inset, we can see a zoom-in on the transition point, where the long-path is  dominant. Hence a DPT takes place. The (black) dashed vertical line represent the transition point.  }
% \end{figure}

% % ===== W values for case 3 ===== %
% \begin{figure}[t]
% \begin{centering}
%  \includegraphics[width=1\columnwidth]{W_at_VB.png}
% \par\end{centering}
% \caption{\label{fig:W case 3} 
% (Color online).
% The (red) circles correspond to the action of the staying put solution and the (blue) dots to the action of the long-path. We can see a  transition point, where the long-path is clearly dominant. The (black) dashed vertical line represent the transition point.}
% \end{figure}

%
%

% ======================================================================================== %

% === Bibliography === %

\bibliographystyle{apsrev4-1}   % the bibliography style; no article names
\bibliography{refs1} % the bibliography file

\end{document}